\documentclass[12pt,preprint]{aastex}
\usepackage[onecolumn,numberedappendix]{emulateapj5}
\usepackage{epsfig}
\usepackage{rotating}
\tightenlines
\newcommand \lsim{\mathrel{\rlap{\lower4pt\hbox{\hskip1pt$\sim$}}
    \raise1pt\hbox{$<$}}}
\newcommand \gsim{\mathrel{\rlap{\lower4pt\hbox{\hskip1pt$\sim$}}
    \raise1pt\hbox{$>$}}}

\newcommand     \kms    {\,{\rm km~s}^{-1}}

\newcommand     \s              {\,{\rm s}}

\newcommand{\sm}{{M_\odot}}

\newcommand{\beq}{\begin{equation}}
\newcommand{\eeq}{\end{equation}}
\newcommand{\beqa}{\begin{eqnarray}}
\newcommand{\eeqa}{\end{eqnarray}}

\newlength{\figwidth}
\countdef\decade=200
\advance\decade by \year
\countdef\hours=201
\advance\hours by \time
\divide\hours by 60
\countdef\mins=202
\advance\mins by \hours
\multiply\mins by 60
\multiply\hours by 100
\countdef\miltime=203
\advance\miltime by \hours
\advance\miltime by \time
\advance\miltime by -\mins

\begin{document}

\title{Astrometry of the Dynamical Ejection of the Becklin-Neugebauer Object from $\theta^1$ Ori C}

%\centerline{DRAFT: \today}

\author{Jonathan C. Tan}

\affil{Dept. of Astronomy, University of Florida, Gainesville, Florida 32611, USA\\jt@astro.ufl.edu}

\begin{abstract}
We show that the proper motion of the Becklin-Neugebauer (BN) object
is consistent with its dynamical ejection from the $\theta^1$~Ori~C
binary, contrary to recent claims by G\'omez et al. Continued radio
observations of BN and future precise astrometric observations of
$\theta^1$~Ori~C with SIM and the Orion Nebula Cluster with GAIA can
constrain the properties of this ejection event, with implications for
theories of how the nearest example of massive star formation is
proceeding.
\end{abstract}


\section{Introduction}\label{S:intro}

Understanding massive star formation remains one of the most
challenging and important problems of contemporary astrophysics
(Beuther et al. 2007; Zinnecker \& Yorke 2007). The complexity of the
process means that massive star formation theories, such as the
turbulent core model (McKee \& Tan 2003), the competitive accretion
model (Bonnell \& Bate 2006) and stellar coalescence model (Bonnell et
al. 1998; Clarke \& Bonnell 2008) require close testing against
observed systems. The closest forming (i.e. accreting) massive star is
thought to be radio source I (Menten \& Reid 1995) within the Orion
Nebula Cluster (ONC), at a distance of $414\pm7$~pc (Menten et
al. 2007, adopted throughout), in the Kleinmann-Low (KL) region. As
reviewed by Tan (2008), this source has been used as observational
evidence in support of all three of the above theories. Part of this
confusion is due to the Becklin-Neugebauer (BN) object, 9.9\arcsec to
the NW (Fig.~1), which is a fast moving (radio-ONC-frame proper motion
of $\mu_{\rm BN}=13.2\pm 1.1\:{\rm mas\:yr^{-1}}$, i.e. $v_{\rm
2D,BN}=25.9\pm2.2\:{\rm km\:s^{-1}}$ towards P.A.$_{\rm
BN}=-27^\circ.5\pm4^\circ$, Plambeck et al. 1995; G\'omez et al. 2008)
embedded B star ($L_{\rm BN}=(2.1 - 8.5)\times 10^3L_\odot$, Gezari,
Backman \& Werner 1998, equivalent to a zero age main sequence mass
$m_{\rm BN,zams} = 9.3\pm2.0\sm$). This proper motion implies that BN
has been moving through the KL region and made a close, possibly
coincident, passage with source {\it I} about 500 years ago. Thus to
understand the nearest example of massive star formation, we need to
understand the origin of BN's motion.

Including the $(+21) - (+8) = +13\:{\rm km\:s^{-1}}$ radial velocity
of BN with respect to the ONC mean (Scoville et al. 1983; Walker
1983), the 3D ONC-frame velocity of BN is $v_{\rm 3D,BN}=29\pm3\:{\rm
km\:s^{-1}}$, and its kinetic energy is $E_{\rm BN} =
(8.3\pm2.3)\times 10^{46} (m_{\rm BN}/10\sm)\:{\rm ergs}$. BN is very
likely to have formed somewhere in the ONC and then attained its high
speed by a close interaction with a massive multiple stellar system
followed by dynamical ejection (Poveda, Ruiz \& Allen 1967).

Tan (2004) proposed BN was launched from the $\theta^1$~Ori~C binary
(also shown in Fig.~\ref{fig:bn}), since this is the only stellar
system in the ONC known to have all the physical properties required
by this scenario: (1) a location along BN's past trajectory
(\S\ref{S:ast}); (2) an (optical)-ONC-frame proper motion
($\mu_{\theta^1C}=2.3\pm0.2\:{\rm mas\:yr^{-1}}$, van Altena et
al. 1988, i.e. $v_{\rm 2D,\theta^1C} = 4.5\pm0.4\:{\rm km\:s^{-1}}$,
towards $\rm P.A._{\theta^1C}=142^\circ.4\pm4^\circ$) that is in the
opposite direction to BN (the direction to BN from $\theta^1$~Ori~C is
a P.A.$=-30^\circ.949$) and is of the appropriate magnitude (the
dynamical mass of BN implied by this motion agrees with the estimate
of $m_{\rm BN,zams}$ and is $m_{\rm BN,dyn}=8.6\pm1.0\sm$ assuming
negligible error in $m_{\theta^1C}=49.5\sm$ and negligible motion of
the pre-ejection triple system in this direction; a pre-ejection
motion of 0.35~mas/yr along this axis (\S\ref{S:high}) would
contribute an additional $1.5\sm$ uncertainty); (3) primary
($m_{\theta^1C-1}=34\sm$) and secondary ($m_{\theta^1C-2}=15.5\sm$)
masses greater than $m_{\rm BN}$ (Kraus et al. 2007); (4) a semi-major
axis of $a=17.0\pm5.8$~AU (Patience et al. 2008) and thus a total
orbital energy ($E_{\rm tot}=Gm_{\theta^1C-1}m_{\theta^1C-2}/(2a)=
(2.7\pm0.9)\times 10^{47}\:{\rm ergs}$) greater than the sum of BN's
kinetic energy and $\theta^1$~Ori~C's kinetic energy ($1.00\times
10^{46}\:{\rm ergs}$) (see Tan 2008 for a review). Note,
$\theta^1$~Ori~C's recoil in this scenario is large enough to remove
it from the Trapezium region (see Pflamm-Altenburg \& Kroupa 2006 for
theoretical studies of the dynamical decay of Trapezium-like systems)
and may be enough to eject it from the ONC completely, with
implications for the effectiveness of its ionizing feedback on
disrupting the star cluster formation process.

Rodr\'iguez et al. (2005) and Bally \& Zinnecker (2005) proposed BN
was launched from an interaction with radio source {\it I}, which
would require this system to be a massive binary, recoiling
away from any large scale ($\gtrsim 100$~AU) gas that it was
originally accreting.
%(see Fig.~\ref{fig:bn}), thought to be a massive protostar (Tan 2008 and
%references therein). 
G\'omez et al. (2008) used the relative motion to BN with respect to
source {\it I} to claim that BN could not have made a close passage
with $\theta^1$~Ori~C, excluding this possibility at the 5-10~$\sigma$
level.

We show in \S\ref{S:ast} that if BN's motion is considered in the
reference frame of the ONC, then a close (coincident) passage with
$\theta^1$~Ori~C is allowed by the data, which permits the scenario of
dynamical ejection of BN from $\theta^1$~Ori~C. In \S\ref{S:high} we
discuss the potential of future high precision astrometric
measurements to constrain the properties of BN's dynamical ejection,
which then constrain BN's interaction distance with source {\it I},
the mass of source {\it I}, and thus the strength of tidal
perturbations on the massive protostar during this encounter.

\vspace{0.2in}

\section{Astrometry of BN in the Orion Nebula Cluster}\label{S:ast}

\begin{figure}[h]
\begin{center}
\epsfig{
%prop3
        file=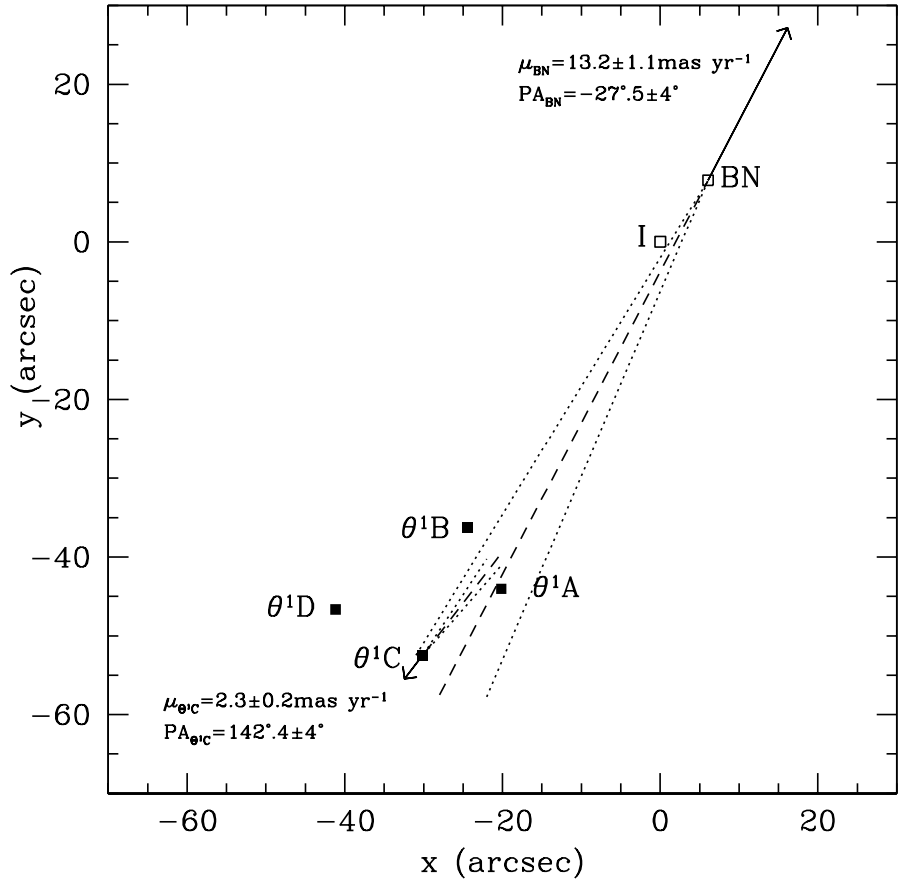,
        angle=0,
        width=\figwidth
}
\end{center}
\caption{ \label{fig:bn} This diagram shows the positions of the
Trapezium stars $\theta^1$~Ori~A, $\theta^1$~Ori~B, $\theta^1$~Ori~C
and $\theta^1$~Ori~D that make up the core of the ONC. The positions
of radio sources I and BN are also shown. The coordinates are relative
to the present position of source I ($\alpha$(J2000)=05 35 14.5141,
$\delta$(J2000)=-05 22 30.556) (Gomez et al. 2008). The proper motions
relative to the cluster of BN (Gomez et al. 2008) and $\theta^1$~Ori~C
(van Altena et al. 1988) are indicated with the arrows. Past
trajectories (dashed line) and $1\sigma$ uncertainties (dotted lines)
are drawn.  }
\end{figure}

To determine BN's past trajectory through the ONC we use the absolute
proper motion of BN ($\mu_\alpha {\rm cos}\delta = -5.3\pm0.9 {\rm
mas\:yr^{-1}}$, $\mu_\delta=9.4\pm1.1 {\rm mas\:yr^{-1}}$ ($1\sigma$
errors); G\'omez et al. 2008) and then correct for the motion of the
ONC (mean of 35 radio sources within central 0.1~pc of ONC:
$\mu_\alpha {\rm cos}\delta = +0.8\pm0.2 {\rm mas\:yr^{-1}}$,
$\mu_\delta=-2.3\pm0.2 {\rm mas\:yr^{-1}}$; G\'omez et al. 2005). The
ONC-frame proper motions are shown in Fig.~\ref{fig:bn}. One sees that
the past trajectory of BN through the ONC overlaps within the
$1\sigma$ errors with the present position of $\theta^1$~Ori~C. Given the
motions of BN and $\theta^1$~Ori~C, the time of coincidence (i.e. when the
dynamical ejection took place) was 4530 years ago, i.e. about 174
orbital periods of $\theta^1$~Ori~C (although the orbital period is only
poorly constrained at present to $26\pm13$~years, Patience et
al. 2008).
%This figure also shows that the
%motion of $\Theta^1C$ relative to the ONC (as defined by the average
%of van Altena et al. 1988's optical sample) is opposite (within the
%uncertainties) to the direction of the vector to BN from $\Theta^1C$.

G\'omez et al. (2008) excluded a coincidence between BN and $\theta^1$~Ori~C
because they used the motion of BN with respect to source {\it I}
(which is measured using relative astrometry to greater accuracy so
has smaller error bars), but did not allow for the fact that their
data indicate that source {\it I} is moving. In the ONC frame this
motion is claimed to be $\mu_\alpha {\rm cos}\delta = -3.7\pm1.2 {\rm
mas\:yr^{-1}}$, $\mu_\delta=-3.4\pm1.3 {\rm mas\:yr^{-1}}$,
corresponding to $\mu_{\rm I}=5.0\pm1.3\:{\rm mas\:yr^{-1}}$
(i.e. $9.9\pm2.6\rm km\s^{-1}$) towards a
P.A.$=+133^\circ\pm16^\circ$. 
%If one measures BN's motion in the ONC
%frame, then, as we have shown, its past trajectory overlaps with
%$\Theta^1C$'s position.

%\section{Motion of the Massive Protostar Source {\it I}}

We note, as a separate point, that source {\it I} is elongated along the
NW-SE axis, i.e. towards P.A.$\simeq+135^\circ$ (Reid et al. 2007). If
the source exhibits variability affecting the location of the centroid
of its emission, then this could lead to an apparent, but false,
proper motion. This effect is a potential source of additional
uncertainty in the motion reported for source {\it I} (and for source
{\it n}) by G\'omez et al. (2008).

Source {\it I} is thought to be a massive protostar and a large proper
motion would be interesting for theories of massive star formation.
F\~ur\'esz et al. (2008) measured the distribution of radial velocities
in the ONC, finding it could be well fit by a Gaussian with
$\sigma_{1D}=3.1\:{\rm km\:s^{-1}}$, for both the entire cluster and
for stars within a 15\arcmin\ radius of the Trapezium. Assuming an
isotropic velocity distribution, the proper motions should exhibit a
Gaussian distribution of motions with $\sigma_{2D}=4.4\kms$. In
comparison, Source {\it I}'s claimed motion of $9.9\pm2.6\rm
km\s^{-1}$ is $(2.3\pm 0.6)\sigma_{2D}$, i.e. not significantly larger
than expected of a typical cluster member. Note, Jones \& Walker
(1988) found $\sigma_{2D}=2.9\kms$ from direct observation of proper
motions (adjusted to $d_{\rm ONC}=414$~pc), for which source {\it I}'s
motion would then be $(3.4\pm 0.9)\sigma_{2D}$. G\'omez et
al. (2005) found $\sigma_{2D}=7.6\kms$ based on proper motions of 35
radio sources, for which source {\it I}'s motion would then be
$(1.3\pm 0.3)\sigma_{2D}$. We conclude, in contrast to G\'omez et
al. (2008), that it is premature to claim that source {\it I} has an
anomalously large motion compared to other ONC stars.

\section{Potential of High Precision Astrometry with SIM}\label{S:high}

For wide angle absolute astrometry, SIM should be able to achieve a
parallax accuracy of about 5~$\mu$as. Assuming a distance of
about 400~pc, this will allow a parallax distance measurement accurate
to 0.2\%, i.e. 0.9~pc. 
%Note the ONC is a couple of parsecs in extent,
%so SIM observations of other stars have the potential to determine the
%3D structure of this cluster, especially if relative astrometry of the
%narrow-angle mode can be used to infer relative distances.

Once the motions of the primary and secondary components of
$\theta^1$~Ori~C due to their binary orbit are accounted for, then the
absolute proper motion of the system should be known to an accuracy of
a few $\mu$as/yr. By averaging over many stars, an even greater
accuracy should be achievable for the absolute proper motion of the
ONC with GAIA. Since $\theta^1$~Ori~C is moving at a few mas/yr in the
ONC frame (van Altena et al. 1988), then the accuracy of the position
angle of the direction of motion would be $\sim0.06^\circ$. Presently it
is only known to about 4$^\circ$.

If, as seems very likely, BN was ejected from $\theta^1$~Ori~C, it should
have been ejected in exactly the opposite direction to $\theta^1$~Ori~C's
motion as measured in the center of mass frame of the pre-ejection
triple system. Comparison of the ONC-frame motion of $\theta^1$~Ori~C with
the present position and ONC-frame motion of BN, will yield
information on motion of the pre-ejection triple system and any
accelerations experienced by the stars since ejection. 

The expected size of pre-ejection triple system proper motion is
uncertain. If the system (with total mass $\simeq 60\sm$) was in
kinetic energy equilibrium with the other ONC stars (with, say,
typical mass $1.0\sm$ and $\sigma_{2D}=4.0\kms$), then we would expect
it to have a plane of sky motion $\sim 0.52\kms$ equivalent to a
proper motion of 0.26~mas/yr. The observed proper motion dispersion of
bright ($V\lesssim12.5$), i.e. massive, stars is $0.70\pm0.06$~mas/yr
(van Altena et al. 1988). Assuming a 0.5~mas/yr proper motion for the
pre-ejection triple system, of which 0.35~mas/yr would be expected to
be tangential to the ejection axis, implies that the ONC-frame proper
motion vectors of $\theta^1$~Ori~C and BN would be misaligned by $10^\circ$
from direct opposition. The current observed misalignment is
$10^\circ\pm6^\circ$. Thus, in the limit that subsequent accelerations
are negligible, high precision ONC-frame proper motions of $\theta^1$~Ori~C
and BN (the latter expected from continued radio observations) can
constrain the motion of the pre-ejection triple system.

%Equivalently an analysis of absolute proper motions of BN and
%$\Theta^1C$ combined with an accurate measurement of the ONC's proper
%motion (e.g. provided by GAIA), will 

The expected gravitational accelerations of $\theta^1$~Ori~C and BN depend
on the distribution of mass in their surroundings. Their trajectories
are taking them away from the ONC center, so they will be experiencing
a deceleration associated with climbing out of the cluster
potential. This effect is largest for BN, but it is still small. BN
has moved 0.12~pc (projected) from the ejection site, and if the
enclosed mass is 500~$\sm$ (likely to be a conservative upper limit,
e.g. Hillenbrand \& Hartmann 1998), then for a starting velocity of
$30\kms$, it would have decelerated by only 0.6~$\kms$.

Close passage with individual stars can also cause more significant
accelerations. $\theta^1$~Ori~C's trajectory may have brought it into
relatively close proximity with $\theta^1$~Ori~A (a B0 star, i.e. $16\sm$,
13\arcsec\ to the NW, with a visual companion at 100~AU of $4\sm$ and
a spectroscopic companion at $\sim 1$~AU of $\sim 3\sm$, Schertl et
al. 2003). However, the relative motion of these stars is only about
1.2~mas~$\rm yr^{-1}$ (van Altena et al. 1988) so that the time of
closest approach would have been about $10^4$~yr ago, long before the
proposed interaction of $\theta^1$~Ori~C with BN.

More importantly, BN made a close passage to source {\it I} about 500
years ago. From the bolometric luminosity of the KL region, source
{\it I} is estimated to have a protostellar mass of about
$20\:M_\odot$.  As an example of the magnitude of the deflections that
can be expected, treating BN as a massless test particle, its
deflection angle due to source {\it I} is $2.25^\circ
(m_{I,*}/20M_\odot)(b/1000{\rm AU})^{-1}(v_{\rm BN}/30{\rm
km\:s^{-1}})^{-2}$, where $b$ is the initial impact parameter and
$v_{\rm BN}$ is the velocity of BN relative to source {\it I}. A
direct trajectory from $\theta^1$~Ori~C's present position (ideally this
would be measured from $\theta^1$~Ori~C's position at the time of ejection)
to BN's position has a closest projected separation from source {\it
I}'s present position of 1.5\arcsec (about 600~AU). Thus an accurate
astrometric solution of this system presents us with the unique
opportunity of constraining the dynamical mass of source {\it I}, the
nearest massive protostar, in combination with the true (unprojected)
distance of closest approach. The true distance of closest approach is
important for evaluating the tidal effects of BN on source {\it I}'s
accretion disk, which are likely to have enhanced accretion to the
star (Ostriker 1995; Moeckel \& Bally 2006). Such enhanced accretion
is likely to have led to enhanced protostellar outflow activity, thus
explaining the $\sim 1000$~yr timescale of the ``explosive'' outflow
from this region (Allen \& Burton 1993; Tan 2004).

\section{Conclusions}

We have reviewed the latest evidence that BN was dynamically ejected
from the $\theta^1$~Ori~C binary, finding that $\theta^1$~Ori~C has all the
physical properties expected in this scenario. We showed that the
trajectory of BN is also consistent with this scenario, in contrast to
recent claims by G\'omez et al. (2008). We discussed how high
precision astrometry of $\theta^1$~Ori~C with SIM can yield information on
the pre-ejection velocity of the system and the size of any subsequent
deflections, in particular that of BN caused by close passage with
source {\it I}, the nearest massive protostar.

\acknowledgements JCT acknowledges support from NSF CAREER grant
AST-0645412 and a grant from NASA for SIM Science Studies.

%--------------------------------------------------------------------------
\end{document}